\documentclass[aps,prl,twocolumn,floats,amsmath,amssymb,graphicx,euscript,superscriptaddress,showpacs]{revtex4}
\usepackage{graphicx}

\newlength{\intwidth}

\def\lapprox{\,\raise0.4ex\hbox{$<$}\kern-0.8em\lower0.7ex\hbox{$\sim$}\,}
\def\gapprox{\,\raise0.4ex\hbox{$>$}\kern-0.8em\lower0.7ex\hbox{$\sim$}\,}
\def\lg{\,\raise0.5ex\hbox{\footnotesize $<$}\kern-0.8em\lower0.5ex\hbox{\footnotesize  $>$}\,}
\def\gl{\,\raise0.5ex\hbox{\footnotesize $>$}\kern-0.8em\lower0.5ex\hbox{\footnotesize  $<$}\,}

\def\be{\begin{equation}}
\def\ee{\end{equation}}
\def\ba{\begin{eqnarray}}
\def\ea{\end{eqnarray}}

\def\bc{\begin{center}}
\def\ec{\end{center}}

\begin{document}
\bibliographystyle{prsty}
\title{Extremely slow spin relaxation in a spin-unpolarized quantum Hall system}

\author{S. Dickmann}
\affiliation{Institute for Solid State Physics of RAS, Chernogolovka 142432, Moscow
District, Russia}

\date{\today}

\begin{abstract}
{Cyclotron spin-flip excitation in a $\nu\!=\!2$ quantum Hall system, being separated from the ground state by a slightly smaller gap than the cyclotron energy and from upper magnetoplasma excitation by the Coulomb gap \cite{di05,ku05}, cannot relax in a purely electronic way but only with the emission of a short-wave acoustic phonon ($k\!\sim\!3\!\cdot\!10^7/$cm). As a result, relaxation in a modern wide-thickness quantum well occurs very slowly. We calculate the characteristic relaxation time to be $\sim\!1\,$s. Extremely slow relaxation should allow the production of a considerable density of zero-momenta cyclotron spin-flip excitations in a very small phase volume, thus forming a highly coherent ensemble -- the Bose-Einstein condensate. The condensate state can be controlled by short optical pulses ($\lesssim 1\,\mu$s), switching it on and off. }
\end{abstract}

\pacs{73.21.Fg, 73.43.Lp, 78.67.De}

\maketitle

\bibliographystyle{prsty}

\vspace{-4mm}
Particular interest in the problem of excitation life-times in low-dimensional electron systems gained popularity in the mid-1990s, when first ideas of possible physical realizations of quantum computations were proposed. Indeed, it is known that one of the requirements (necessary though insufficient!) is the long lifetime of an excited quantum state in a two-level system, considered as a qubit. Most of those works were devoted to zero-dimensional objects -- the electron relaxation in quantum dots (see \cite{amasha} and references therein) or to a one-dimensional system -- quantum Hall edge electrons \cite{1D}. The relaxation in strongly correlated two-dimensional systems was studied mainly in particular cases of spin relaxation in a spin-polarized quantum Hall system (i.e. in a quantum Hall ferromagnet \cite{fu08,di-zi}).
\vskip -.5mm
In this letter we report on the relaxation rate calculation of the {\it cyclotron spin-flip exciton} (CSFE) in an {\it unpolarized} even-integer quantum Hall system, where an electron is effectively promoted from the fully occupied Landau level to the next fully empty level with a spin flip \cite{di05,ku05} (at which point an {\it effective hole} appears in the initial level, see Fig. 1). The CSFE with spin numbers $S\!=\!S_z\!=\!1$ is the {\it lowest energy excitation} in the system (Fig. 1). Yet it is separated from the ground
state by a wide gap, which is only somewhat smaller than the cyclotron one. { We will see that due to a concatenation of circumstances -- a specific combination of certain CSFE features, the relaxation rate should be very slow. One of the features is the optical inactivity of the CSFE, which takes place even in the presence of spin-orbit interaction \cite{foot}.}
Hence the relaxation could only occur non-radiatively due to small perturbations violating the spin and energy conservations: spin-orbit and electron-phonon
couplings.
We estimate that the CSFE life-time is of the order of one second. Thus the predicted relaxation time could compete with, or even be longer than, the record spin relaxation time $T_1\sim 0.1\!-\!1\,$s, fixed experimentally at low magnetic fields in a single-electron quantum dot \cite{amasha}.

The studied CSFEs may represent a highly coherent ensemble -- a Bose-Einstein condensate (c.f. \cite{foot1}) -- where the number of excitons could be macroscopically large -- the exciton concentration actually reaches several percent of the total number of electrons.
Due to this, the relaxation of the CSFE ensemble occurs nonexponentially. Meanwhile it can be
created and quenched (switched on/off) by means of short laser pulses, so the real time of the
switching on/off is estimated to be $\lesssim\!1\,\mu$s. Thus the CSFE ensemble should be optically controllable.

\begin{figure}[h]
\vspace{-10.mm}
\begin{center}
\includegraphics*[angle=-90,totalheight=80.mm,width=.79\textwidth]{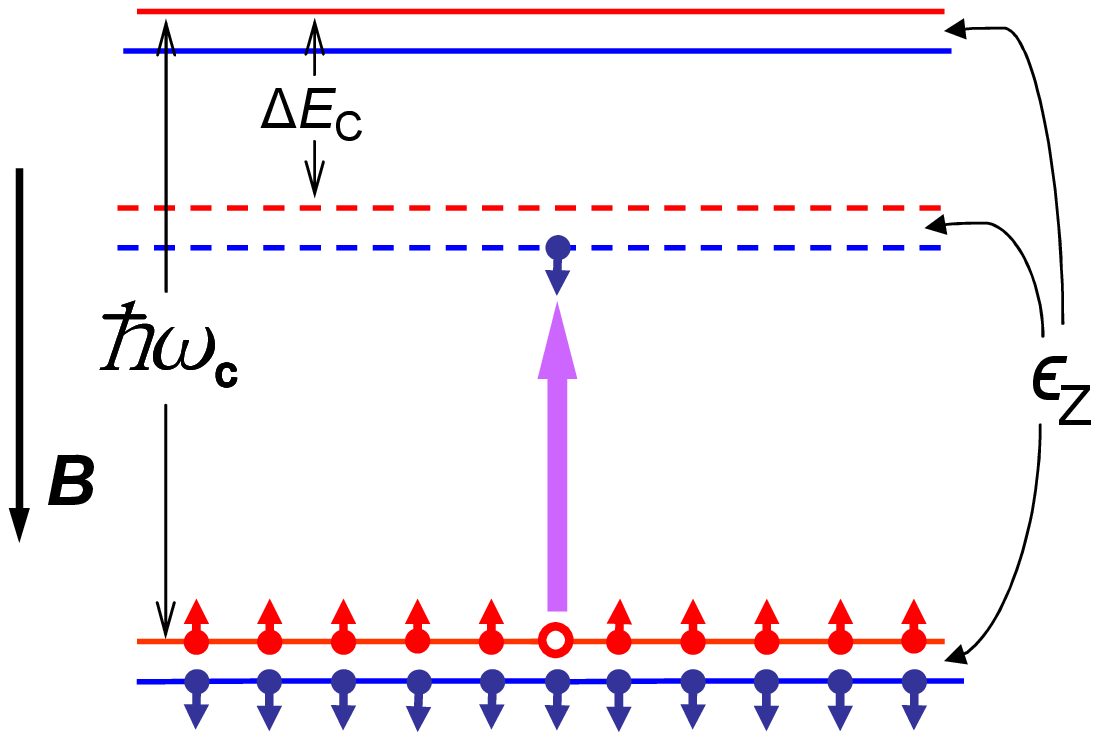}
\end{center}
\vskip -47.mm
\caption{\vspace{-0.mm}
Illustrating the $S_z\!=\!1$ cyclotron spin-flip exciton at the $\nu\!=\!2$ filling. (The magnetic field is assumed to be directed downwards.) Thick upward arrow shows the effective promotion of an electron, in the limit $q\!\to\!0$, from the zero Landau level to the first with a spin flip. The cyclotron and Zeeman gaps, $\hbar\omega_c$ and $\epsilon_{\rm Z}$, as well as the Coulomb shift, $\Delta E_{\rm C}$, are shown. See Eq. (2).
\vspace{-1.mm}}
\end{figure}

{\it Excited state} --- The CSFE spectrum in the $\nu\!=\!2$ case, studied earlier both theoretically \cite{di05,ka84} and experimentally \cite{ku05}, represents a triplet with the $S\!=\!1$ and $S_z\!\!=\!0,\pm 1$ spin components separated by the Zeeman gap. By analogy with previous works devoted to the calculation of excitations spectra in purely electronic quantum Hall systems (e.g. see \cite{di05}) as well as to
excitation relaxation \cite{di-zi}, we employ the {\it excitonic representation} technique, where {\it exciton states} are used as a basis set instead of single electron Fermi states. The exciton states are generated by exciton creation operators \cite{di05,ku05,di-zi,dz83},
 \begin{equation}\label{Q}
  {\cal Q}_{ab\,{\bf q}}^{\dag}=\frac{1}{\sqrt{ {\cal N}_{\phi}}}\sum_{p}\,
  e^{-iq_x p}
  b_{p+\frac{q_y}{2}}^{\dag}\,a_{p-\frac{q_y}{2}}\,,
\end{equation}
\vskip -2.mm
\noindent acting on the ground state $|0\rangle$. Here $a_p$ and $b_p$ are electron Fermi annihilation operators corresponding to the``initial'' and ``final'' states of the promoted electron, and ${\bf q}$ is the exciton dimensionless wave vector. Index $p$ labels intrinsic Landau level states $\psi_{np}({\bf r})\!=\!(2\pi {\cal
N}_\phi)^{-1/4}e^{ipy}\varphi_n(p\!+\!x)$, where
$\varphi_n(x)$ is the oscillator wave function, $n$ is the Landau level number, and ${\cal N}_\phi$ is the number of magnetic flux quanta (we measure length in units of the magnetic length $l_B$). $a$ and $b$ are binary indexes indicating both
the Landau level number and the spin state. For example, for the $S_z=1$ component of the
CSFE at $\nu\!=\!2$, these are $a\!=\!(0,\downarrow)\equiv\overline{0}$ and
$b\!=\!(1,\uparrow)\equiv{1}$, see Fig. 1 ($n$ and $\overline{n}$ are spin-up
and spin-down sublevels).

The set of exciton operators (1) with arbitrary indexes $a$ and $b$ obey a closed Lie algebra. The basic property of the exciton states, ${\cal Q}_{ab\,{\bf
q}}^{\dag}|0\rangle$ consists in the fact that they diagonalize a considerable part in the exact many-electron Coulomb interaction Hamiltonian $H_{\rm int}$.
At $\nu\!=\!2$ the excitonic representation of the ${\bf q}\!=\!0$ CSFE state,
${\cal Q}^\dag_{\overline{0}1,{\bf 0}}|0\rangle$, allows the determination of first order corrections to the state in terms of the interaction $H_{\rm int}$  and thus to calculate the CSFE energy perturbatively up to the second order in terms of $H_{\rm int}$. Precisely this second-order result yields in this case the leading contribution to the excitation Coulomb energy \cite{di05}.

On the basis of the results \cite{di05,ku05,ka84} one can conclude that in the leading approximation in small parameters $q$ and
$r_{\rm s}=\!\alpha(e^2/\kappa
l_B)/\hbar\omega_c$ ($\kappa$ is the GaAs dielectric constant, and $\alpha\!<\!1$ is the averaged form-factor arising due to the finiteness of the 2D electron-layer thickness) the CSFE spectrum is
\begin{equation}\label{energy}
  E(S_z, { q})=\hbar\omega_c -\epsilon_{\rm Z}S_z-\Delta E_{\rm C}+q^2/2M_{\rm x}\,,
\end{equation}
where the negative value $-\Delta E_{\rm C}$ is the Coulomb shift calculated to the second order in $r_{\rm s}$ and therefore expressed in units of 2Ry$=\!\!(e^2/\kappa l_B)^2/\hbar\omega_c
\!=\!m_e^*e^4/\kappa^2\hbar^2$ \cite{di05}, whereas $\!1/M_{\rm x}$
is a coefficient calculated to the first order in $r_{\rm s}$ \cite{ka84}.
 $\Delta E_{\rm C}$ is definitely a positive value \cite{di05}, whereas
the sign of $\!1/M_{\rm x}\!=\!\!(e^2/\kappa l_B)\!\int_0^{\infty}\!\!{dp}F(p)
e^{-p^2/2}(p^2\!/2\!-\!p^4\!/4)$ varies with a finite thickness form-factor $F(q)\!=\!\int\!\!\int
  dz_1dz_2e^{-q|z_1-z_2|/l_B}|\chi(z_1)|^2|\chi(z_2)|^2$ ($\chi$ is the
size-quantized wave-function of an electron confined in the $z$-direction). In the ideal 2D case, $F\!=\!1$  and $1/M_{\rm x}$ is negative (though
at $q\gapprox 1\;\:$ $E(S_z,q)$ increases with $q$ \cite{ka84}). In the case of modern wide quantum-well structures $F(q)$ rapidly decreases with $q$, and $1/M_{\rm x}$ becomes positive. We will consider precisely this real situation, i.e. assume that $1/M_{\rm x}\!>\!0$. Our final result does not depend actually on $1/M_{\rm x}$.
In practice the $\Delta E_{\rm C}$ and $1/M_{\rm x}$ values can be found from experimental data. According to \cite{ku05} one obtains $\Delta E_{\rm
C}\approx 0.35\,$meV. There are no direct measurements yielding $1/M_{\rm x}$, but we can estimate this and find that $\Delta E_{\rm
C}\!<\!1/M_{\rm x}\!\leq\! 1\,$meV, so that $1/M_{\rm x}\!\gg\!T\!\sim\!0.1\,$K. Since $\epsilon_{\rm Z}=0.0255B\,$meV and $\hbar\omega_c=1.73B\,$meV
($B$ is in Teslas), hence the first term in Eq. \eqref{energy} is always much larger than all the remaining terms.

{\it Relaxation} --- How can the CSFE decay? This process is determined by two necessary conditions: by
the availability of an interaction that does not conserve the electron system spin, and by a mechanism of energy dissipation making the relaxation process irreversible. If comparing to possible mechanisms of the spin-exciton relaxation in a quantum Hall ferromagnet \cite{di-zi}, the analysis
shows that in our case the only CSFE relaxation channel is governed by the spin-orbit and electron-phonon couplings, leading to acoustic phonon emission. The decay probability is determined by the Fermi golden rule, \vspace{-3.mm}
\begin{equation}\nonumber
w_{fi}=(2\pi/\hbar)|{\cal M}_{fi}|^2\delta(E_f-E_i), \vspace{-1.5mm}
\end{equation}
where in the initial state $|i\rangle$ the number of SCFEs is greater by one than in the final $|f\rangle$, and conversely the number of phonons is greater by one in $|f\rangle$ than in $|i\rangle$, ${\cal M}_{fi}$ being the relevant matrix element.

Methodically it is useful to develop the approach as applied to the general case where $\nu\!=\!2n\!+\!2$  ($n=0,1,2,...$). The spin-orbit coupling is described by a single electron term of the total Hamiltonian, namely
${\hat H}_{\rm so}\!=\!\alpha\left(\hat{{\bf q}}\times\hat{\mbox{\boldmath $\sigma$}}
\right)_{\!z}\!+\!
\beta\left(\vphantom{\left(\hat{{\bf q}}\times\hat{\mbox{\boldmath
  $\sigma$}}
  \right)}\hat{ q}_y\hat{\sigma}_y\!-\!{\hat q}_x\hat{\sigma}_x\right)$, where $\hat{{\bf q}}\!=\!-i{\bf\nabla}\!+\!e{\bf A}/c\hbar$.  This operator represents a combination of the Rashba ($\sim\alpha$) and Dresselhaus ($\sim\beta$) terms \cite{by84} and does not violate translational symmetry. As usual \cite{di-zi} it is convenient to employ a bare single-electron basis diagonalizing the single-electron Hamiltonian $\hat{\bf q}^2/2m_e^*+{\hat H}_{\rm
so}$. Within the leading order in
the ${\hat H}_{\rm so}$ terms one obtains the basis states,
\vspace{-3.mm}
\begin{equation}\label{so_ab}
\begin{array}{l}
\Psi_{np}=
  \left( {\psi_{n p}\atop v\sqrt{n\!+\!1}\psi_{n\!+\!1\,p}
  +iu\sqrt{n}\psi_{n\!-\!1\,p}}\right)\;\mbox{and}{}\qquad\quad\quad\vspace{1.mm}\\
  \qquad\qquad\Psi_{\overline{n}p}=\left({-v\sqrt{n}
  \psi_{n\!-\!1\,p}
  +iu\sqrt{n\!+\!1}\psi_{n\!+\!1\,p}\atop
  \psi_{n p}}\right),\!\!
\end{array}
\vspace{-3.mm}
\end{equation}
where $u=\beta\sqrt{2}/l_B\hbar\omega_c$ and
$v=\alpha\sqrt{2}/l_B\hbar\omega_c$ are small dimensionless parameters.
Thus the single-electron states acquire a chirality labeled by subindex $n$ or ${\overline n}$
instead of pure spin-up ($b$) and spin-down ($a$) quantum numbers.  The definition of the exciton
creation operator formally remains the same [Eq. (\ref{Q})], although the $a_p$ and $b_p$
operators now describe annihilation in some states (\ref{so_ab}).  In  particular, in the case of
the CSFE $a_p$ corresponds to annihilation in the state $\Psi_{\overline{n}p}$ state and
$b_p^\dag$ to creation, resulting in the $\Psi_{n\!+\!1p}$. Exactly this transition from
$\Psi_{\overline{n}p}$ to $\Psi_{n\!+\!1p}$
now represents the $\overline{n}\!\to\!{n\!+\!1}$ promotion, and our task is to calculate the
relaxation of the ${\cal
Q}^\dag_{\overline{n}{n\!+\!1}}{\bf q}|0\rangle$ state.

The Hamiltonian of electron coupling to 3D acoustic phonons with momenta ${\bf k}\!=\!({\bf q},k_z)$ is written as
\begin{equation}\label{e-ph1}
{\hat H}_{\rm e-ph}\!=\!\frac{{\hbar}^{1/2}}{L L_z^{1/2}}
  \!\sum_{{\bf q},{k}_z,s}\!
  {U'}_{s}({\bf k})\,{\hat P}_{{\bf k},s}^\dag
  {\cal H}_{{\rm e-ph}}({\bf q})+
  \mbox{H.c.}
\end{equation}
\vskip -1.5mm
\noindent
(e.g., see Ref. \cite{di-zi}), $L^2\!=\!2\pi {\cal N}_\phi l_B^2$ is the 2D area, $L_z$ is the thickness of the slab, ${\hat P}_{{\bf k},s}^\dag$ is the phonon creation Bose operator, i.e.  state ${\hat P}_{{\bf k},s}^\dag|0\rangle$ represents a combination of the 2D electron ground  state and a 3D phonon with momentum ${\bf k}$ and polarization $s$), ${\cal H}_{\rm e-ph}({\bf q})\!=\!\int\!e^{i{\bf qr}}
   {\hat \Psi}^{\dag}({\bf r}){\hat \Psi}({\bf r})d^{2}r$, and ${U'}_{s}({\bf k})$ is the renormalized vertex, ${
U'}_{s}({\bf k})\!=\!\!U_{s}({\bf k})\Phi(k_z)$, where\vspace{-2.mm}
\begin{equation}\label{FF}
\qquad\Phi(k_z)\!=\!\int\!e^{ik_{z}z}|\chi(z)|^2dz\,.\vspace{-1.mm}
\end{equation}
\vskip -2.mm
\noindent When substituting ${\hat \Psi}({\bf r})\!=\!\sum_{np}\left[a_{np}\Psi_{np}\!+\!b_{np}\Psi_{\overline{n}p}\right]\!$, we only keep in ${\cal H}_{\rm e-ph}$ the terms governing the decay of the ${\cal Q}^\dag_{\overline{n}{n\!+\!1}{\bf
q}}|0\rangle$ state, and the dimensionless operator ${\cal H}_{\rm e-ph}$ relevant to the case takes the form
${\cal H}_{\rm e-ph}({\bf q})\!\!=\!\!\sqrt{{\cal N}_{\phi}}G_n(q){\cal Q}_{\overline{n}{n\!+\!1}\,{\bf q}}\!$ where $G_n(q)\!=\!L_n^1(q^2/2)e^{-q^2/4}\left[v(q_+)^2\!+\!iuq^2/2\right]/(n\!+\!1)\:$
[$q_+\!\!=\!\!-i(q_x\!+\!iq_y)\!/\!\sqrt{2},$ $L_n^1$ is the
Laguerre polynomial].

Further manipulations are simplified in view of a basic feature of the studied relaxation
process -- only `hard' phonons (with frequency~$\!\approx\!\omega_c)$ ) are generated at the CSFE annihilation.
Besides, the calculation shows that only phonons emitted almost parallel to the ${\hat z}$-direction
are relevant, i.e. $q\lesssim\!1\ll\!k_z\!\approx\! l_B\omega_c/c_s$. In this connection, there are also
apparent simplifications for the vertex $U_{s}$, namely: (i)  only the contribution of the deformation phonon field where the amplitude is proportional to $\sqrt{k}$ has to be taken into account, and (ii) only LA phonons ($s\!=\!l$) give rise to the deformation potential in the GaAs lattice \cite{gale87}. For the 3D vertex one only needs the expression for the square (c.f. \cite{di-zi}): $|U_l|^2=\pi\varepsilon_{\rm ph}({k_z})/p_0^3\tau_{D}$, where the phonon energy is considered to be proportional to $k$:
$\varepsilon_{\rm ph}(k)=\hbar c_l\sqrt{k_z^2\!+\!q^2}/l_B$. $p_0=2.52\!\cdot\!
10^6\,$cm${}^{-1}$, $\tau_D\simeq 0.8\,$ps (see \cite{di-zi,gale87}) and $c_l\approx 4.5\!\cdot\! 10^5\,$cm/s \cite{dorner90} are the material parameters of GaAs.

To calculate the transition matrix element  ${\cal M}_{fi}$ one has to choose the initial and final
states. In principle, those may correspond to a single exciton decay: $|i\rangle\!=\!{\cal
Q}^\dag_{\overline{n}{n\!+\!1}\,{\bf
q}_0}|0\rangle\!\to\!|f\rangle\!=\!{\hat P}_{{\bf k},l}|0\rangle$, or to an exciton-exciton scattering process: $|i\rangle\!\!=\!\!{\cal Q}^\dag_{\overline{n}\,{n\!+\!1}\,{\bf
q}_1}\!{\cal Q}^\dag_{\overline{n}\,{n\!+\!1}\,{\bf
q}_2}|0\rangle\!\to\!|f\rangle\!\!=\!\!{\hat P}_{{\bf k},l}{\cal
Q}^\dag_{\overline{n}\,{n\!+\!1}\,{\bf q}'}|0\rangle$.
Respectively, the matrix element is equal to
 ${\cal M}_{fi}^{(1)}(k_z,{\bf q}_0)\!\!=\!\!\frac{1}{l_B}\sqrt{\frac{\hbar}{2\pi L_z}}\sum\limits_{{\bf q}}U_l(k_z)\Phi(k_z)G_n(q)M_1({\bf q},{\bf q}_0),\,$ or to
\vskip -4.mm
\begin{equation}\begin{array}{l}
{\cal M}_{fi}^{(2)}(k_z,{\bf q}',{\bf q}_1,{\bf q}_2)\nonumber\\=\frac{1}{l_B}\sqrt{\frac{\hbar}{2\pi L_z}}\sum\limits_{{\bf q}}U_l(k_z)\Phi(k_z)G_n(q)M_2({\bf q}',{\bf q};{\bf q}_1,{\bf q}_2),\nonumber
\end{array}
\end{equation}
\vspace{-3.ex}

\noindent where ${}\!M_1(...)\!\!=\!\!\left\langle 0|{\cal Q}_{{{\bf q}}}{\cal
Q}_{{\bf q}_{0}}^{\dag}|\,0\right\rangle\!\equiv\! \delta_{{\bf q},{\bf q}_0}$ and~$M_2(...)=\left\langle 0|{\cal Q}_{{{\bf q}}'}{\cal Q}_{{{\bf q}}}{\cal
Q}_{{\bf q}_{\!1}}^{\dag}\!{\cal Q}_{{\bf
q}_2}^{\dag}|\,0\right\rangle$ (${\cal Q}_{\bf q}$ actually stands for ${\cal
Q}_{\overline{n}\,n\!+\!1\,{\bf q}}$). The latter expectation is calculated with the help of excitonic commutation algebra \cite{di05,di-zi}:\vspace{-3.mm}
\begin{equation}\label{expectations}
{}\!\!\!M_2\!=\!\displaystyle{\delta_{{\bf q}',{\bf q}_2}\delta_{{\bf q},{\bf q}_1}\!\!+\!\delta_{{\bf q}',{\bf q}_1}\delta_{{\bf q},{\bf q}_2}\!-\!\frac{2\cos{\!\phi}}{{\cal N}_\phi}\delta_{{\bf q}_1\!+{\bf
q}_2,\,{{\bf q}}'\!+{{\bf q}}}},\!
\end{equation}
\vskip -4.mm
\noindent where $\phi=\left({{\bf q}}'\!\times\!{{\bf q}_{{}\,2}}\!+\!{{\bf q}}\!\times\!{{\bf q}_1}\right)_z\!/2$.

Generally, in order to calculate the CSFE relaxation rate, $R\!=\!\sum_{if}w_{if}$, one should know the distribution $N_{\bf q}$ of excitons over the {\bf q} wave numbers.  At
any moment the CSFE distribution is quasi-equilibrium one, and characterized
by an `adiabatic' chemical potential, since the thermodynamic equilibrium is
certainly established much faster than the CSFE decay processes
occur. (The excitons obey Bose
statistics because their number in any state determined by a
certain ${\bf q}$ may be macroscopically large.) Our conditions are as follows:
(i) initially the total number of excitons $\,{\cal N}_{\rm
x}\!=\!\sum_{\bf q}\!N_{\bf q}$
excited by a short external optical pulse is rather large: ${\cal N}_{\rm x}\!\sim\!0.01{\cal N}_\phi$; (ii) the relevant values are $q\sim\sqrt{M_{\rm x}T}\ll 1$, and
$N_{\bf q}\!\sim\!e^{-q^2/2M_{\rm x}T}$ if $q^2\!\gtrsim\!M_{\rm x}T$. First, we estimate the single-exciton relaxation rate:\vspace{-2.mm}
\begin{eqnarray}\nonumber\vspace{-10.mm}
   R_1=\displaystyle{\frac{2\pi}{\hbar}\sum_{k_z,{\bf q}_0}|{\cal M}_{fi}^{(1)}({\bf q}_0)|^2
\delta(\hbar\omega_c\!-\!\hbar c_l\sqrt{k_z^2\!+\!q_0^2}\,/l_B\,)}\vspace{-5.mm}\nonumber\\\displaystyle{{}\!\!\approx
\frac{|U_l(\omega_c/c_l)\Phi(\omega_c/c_l)|^2}{\pi\hbar l_B^2 c_l}\!\sum_{{\bf q}_0}\!N_{{\bf q}_0}|G_n(q_0)|^2\sim {\cal N}_{\rm x}/\tau_1}\,,\nonumber
\end{eqnarray}
\vskip -3.mm
\noindent where $1/\tau_1\!=\!(M_{\rm x}T)^2/\tau$ -- see $\tau$ below defined in Eq. (7). When calculating the rate due to the two-exciton scattering, one finds that under our condition
the dominant contribution to the rate is provided by the $\sim\!1/{\cal N}_\phi$ term of the expectation (\ref{expectations})  (also $\phi\approx 0$ has to be set), and the rate is\vspace{-2.mm}
\begin{eqnarray}\nonumber\vspace{-4.mm}
  R_2=\displaystyle{\frac{\pi}{\hbar}}\,
  {\sum_{k_z,{\bf q}',{\bf q}_1,{\bf q}_2}\!\!|{\cal M}^{(2)}_{fi}(k_z,{\bf q}',{\bf q}_1,{\bf q}_2)|^2}\quad\qquad{}\qquad\qquad\nonumber\\\times\,{N_{{\bf q}_1}\!N_{{\bf q}_2}
\delta(\hbar\omega_c\!-\!\hbar c_l\sqrt{k_z^2\!+\!{q'}^2}/l_B\,)}{}\quad\qquad\nonumber\\\vspace{-2.mm}
\displaystyle{\approx\!
\frac{2|U_l(\omega_c/c_l)\Phi(\omega_c/c_l)|^2}{\pi\hbar l_B^2 c_l{\cal N}_\phi^2}\!\!}\!\! \sum\limits_{{\bf q}_1,{\bf q}_2,{\bf q}'}\!\!\!\!\!\!N_{{\bf q}_1}\!N_{{\bf q}_2}\!\left|G_n(|{\bf q}_1\!+\!{\bf q}_2\!-{\bf q}'|)\right|^2\!.\nonumber
\end{eqnarray}
\vskip -2.mm
\noindent The summation in this expression is reduced to \vspace{.5mm}
$\sum_{{\bf q}_1,{\bf q}_2} \!\!N_{{\bf q}_1}\!N_{{\bf
q}_2}\!\sum_{{\bf q}'}\!\left|G_n({q}')\right|^2\!\!\!\!=\!\!2N_\phi
{\cal N}_{\rm x}^2(u^2\!+\!v^2)$.
Finally, if $n_{\rm x}\!=\!{\cal N}_{\rm x}/{\cal N}_\phi$ is the CSFE concentration, then the kinetic equation takes the form $-dn_{\rm x}/dt=n_{\rm
x}^2/\tau$ with characteristic inverse relaxation time
\begin{equation}\label{tau2}
\frac{1}{\tau}=\frac{4\omega_c|\Phi(\omega_c/c_l)|^2(u^2\!+\!v^2)}{p_0^3l_B^2c_l\tau_D}\,.
\end{equation}
Obeying equation
$n_{\rm x}(t)=\displaystyle{{n_{\rm x}(0)}/({1+n_{\rm x}(0)t/\tau}})\,,$
the SCFE density decays nonexponentially. Yet, note that due to
the nonexponentiality of the relaxation, the real value that
should be compared with the experimental results is not $\tau$ but
$\tau'\!\sim\!\tau/n_{\rm x}(0)$.

Meanwhile the single-exciton relaxation, though exponential, occurs much more slowly -- the characteristic relaxation time $\tau_1$ is by a huge factor $\sim(M_{\rm x}T)^{-2}\!\sim\! 10^{3}\!-\!10^{4}$
longer than in the case of exciton-exciton scattering (7). This
feature is determined by considerable enhancement of the relaxation phase volume in the case of two-exciton scattering processes.

{\it Discussion} --- When numerically estimating $\tau$, one
faces the basic difficulty related to uncertainty in the
form-factor $\Phi(\omega_c/c_l)$ value. This strongly depends on the
poorly observable function $\chi(z)$. However, it is clear that
for a wide quantum well with effective thickness $d\simeq 20\,$nm
and for a magnetic field, e.g. equal to $5\,$T, a considerable
incommensurability of $d$ and $\omega_c/c_l\!=\!k_z\!\simeq\!3/$nm
takes place. This fact tremendously reduces the Fourier component \eqref{FF}
. When estimating $\chi(z)$
by means of three models: (i) Fang and Howard;
(ii) Takada and Uemura (see \cite{ando} and references
therein); and (iii) the simple
model where quantum-well walls are considered to be infinitely
high, one finds that $|\Phi(k)|^2\approx {\cal C}/(kd)^6$ where
${\cal C}\approx 4.7\!\cdot\!10^4,\;6.5\!\cdot\!10^3$, and $3.1\!\cdot\!10^3$
respectively ($d/2$ in all three models is set equal to the
average penetration length of the charge into the semiconductor).
Substitution of this estimate into Eq.
\eqref{tau2} and the assumption that $u^2\!+\!v^2\!=\!10^{-3}/B$ ($B$
is in Teslas) \cite{di-zi} yields for a quantum well with
$d\!=\!20\,$nm: $
    \tau\approx 10^{-2}\!\cdot\! B^5/{\cal C}\,$,
and hence $\tau'\sim B^5/{\cal C}$. At $B\!=\!5\,$T even for the
`fastest' Fang-Howard model describing the $\chi(z)$
wave-function, one gets estimates of $\sim\!7\,$ms and $\sim\!0.7\,$s
for the $\tau$ and $\tau'$ times respectively.

The CSFE was diagnosed in experiment \cite{ku05} as a result of Raman scattering.
However, the direct photo-absorption of photons with angular momenta -1 ($\sigma^-$-photons) in the vicinity of laser carrier frequency
$\omega_{\rm on}\!\approx\!\omega_c^h\!+\!\omega_c\!+\!(E_g\!+\!\epsilon_{\rm Z}/2-3|g_h\mu_BB|/2)/\hbar$ would seem to be more intensive way of CSFE creation [$E_g(B)$ is the GaAs/AlGaAs forbidden gap depending on the magnetic field; $\omega_c^h$ and $g_h$ describe the cyclotron frequency and the $g$-factor of the valence-band heavy-holes]. This pulse creates the $S_z\!=\!1/2$ electrons in the first Landau level and the $S_z\!=\!-3/2$ heavy-holes in the first Landau level within the valence band. The latter, due to some fast relaxation and recombination processes, should convert into the effective $+1/2$ hole in the electronic zeroth Landau level. As a result the CSFE emerges. One could also quench the CSFE by using, e.g. a $\sigma^+$ pulse with the frequency $\sim\!(E_g\!-\!\epsilon_{\rm Z}/2\!+\!3|g_h\mu_BB|/2)/\hbar$ and simultaneously a $\sigma^-$ pulse at the frequency $\!\sim\! (E_g\!+\!\epsilon_{\rm Z}/2\!-\!3|g_h\mu_BB|/2)/\hbar$  filling the zeroth electronic Landau level. As a result one can conclude that even the pulses' widths $\lesssim 1\,\mu$s would be sufficient for exciting/quencing, and the ratio of the excited state lifetime $\tau'$ towards the time of the state switching on/off is greater than $10^6$.

Finally, we concern the relaxation of other excitations with energies higher than but close to the same energy $\hbar\omega_c$. The spinless magnetoplasma mode, described excitonically as $2^{-1/2}\!\!\left({\cal
Q}^\dag_{\overline{0}\overline{1}}\!+\!{\cal
Q}^\dag_{{0}{1}}\right)|0\rangle$, has a considerable radiative relaxation channel through the emission of a photon with energy $\hbar\omega_c$ and therefore decays within $\sim 5\!-\!100\,\mu$s. Analysis shows that, due to spin-orbit coupling and processes of phonon absorption, the rate of an activation `overflowing' the CSFE into the spinless magnetoplasmon constitutes a value $\lesssim 10^8\!\cdot\!e^{-\delta E/T}\!{}$, where $\delta E\!=\!\Delta E_{\rm C}\!+\!\epsilon_{\rm Z}\approx 0.4\,$meV at $B\!=\!5\,$T. Hence at $T\!<0.25\!$\,K the thermal activation time of the CSFE conversion into the magnetoplasma mode is definitely longer than $1\,$s. Two other components of the CSFE triplet are  also `dark' excitons, they do not decay radiatively but, due to spin-orbit coupling and processes of inter-exciton scattering, convert within time $\lesssim 1\,\mu$s into the $S_z\!=\!1$ lowest CSFE mode.

In summation, we have studied nonexponential relaxation of the lowest-energy excitation in the $\nu\!=\!2$ unpolarized quantum Hall system. Our estimates should be valid for wide quantum wells (where the thickness is $>\!20\,$ns), for magnetic fields greater than $5\,$T and temperatures $\lesssim 0.25\,$K. The studied excitation, representing a cyclotron spin-flip mode, should actually be metastable because even the fastest relaxation mechanism results in a characteristic life time in the order of $1\,$sec. As opposed to the relaxation in a quantum dot \cite{amasha}, relaxation of the studied excitation should sharply decelerate with an increasing magnetic field. The metastability of the state is determined by several reasons: (i) the studied relaxation is simultaneously the energy and spin relaxation process (radiative relaxation is forbidden \cite{foot}); (ii) the state is energetically distant from the ground state, so the emitted phonon, possessing a very short wavelength, hardly couples with the state; and (iii) due to a considerably negative Coulomb shift the state is also distant from the upper magnetoplasma mode -- {\it this fact strongly favors metastability actually canceling a thermal-activation transition to a radiatively relaxing state}. The CSFE mode could be controlled optically by exciting and `quenching' pulses with widths $\lesssim 1\,\mu$s. The long-lived excitons have to form an exciton Bose-Einstein condensate. The number of excitons in the condensate is governed by the amplitude and width of the optical pulse exciting the system. This state should be similar to a `thermodynamic condensate' of spin waves in a quantum Hall ferromagnet \cite{di-zi}, where excitons' momenta are not strictly equal to zero due to an inhomogeneity  caused by an external disorder, but belong to a phase volume that is much smaller than the total number of excitons in the condensate (see Ref. \cite{di-zi}). Decay of the condensate should also be characterized by time \eqref{tau2}.

The author acknowledges the support of the Russian Fund of Basic Research and gives thanks to I.V. Kukushkin and L.B. Kulik for many useful discussions.


\begin{thebibliography}{99}
\bibitem{foot}
{ This is true if radiative CSFE decay is calculated within the dipole approximation, however, in principle, the CSFE radiation channel does exist. It is determined by spin-orbit coupling and by the higher degree term of the multi-pole radiation. At  $B\!=\!5\,$T calculation results in $\gtrsim\!10\,$s for the characteristic time of the radiative CSFE relaxation channel.}
\end{thebibliography}
\end{document}